\documentclass[aps,prl,twocolumn,showpacs]{revtex4}
\usepackage{graphicx}

\begin{document}

\title{
Pseudogap, Superconducting Gap, and Fermi Arcs in Underdoped
Cuprate Superconductors }

\author{Hai-Hu Wen$^1$, Lei Shan$^1$, Xiao-Gang Wen$^2$, Yue Wang$^1$, Hong Gao$^1$, Zhi-Yong Liu$^1$, Fang Zhou$^1$, Jiwu Xiong$^1$, Wenxin Ti$^1$}

\affiliation{$^1$National Laboratory for Superconductivity,
Institute of Physics, Chinese Academy of Sciences, P.~O.~Box 603,
Beijing 100080, P.~R.~China}

\affiliation{$^2$Department of Physics, Massachusetts Institute of
Technology, Cambridge, Massachusetts 02139, USA}

\date{\today}

\begin{abstract}
Through the measurements of magnetic field dependence of specific
heat in $La_{2-x}Sr_xCuO_4$ in zero temperature limit, we find
that the nodal slope $v_\Delta$ of the superconducting gap has a
very similar doping dependence of the pseudogap temperature $T^*$
or value $\Delta_p$. Meanwhile the maximum quasiparticle gap
derived from $v_\Delta$ is quite close to $T^*$. Both indicate a
close relationship between the pseudogap and superconductivity. It
is also found that $T_c \approx \beta v_\Delta \gamma_n(0)$, where
$\gamma_n(0)$ is the extracted zero temperature value of the
normal state specific heat coefficient which is proportional to
the size of the residual Fermi arc $k_{arc}$. These observations
mimic the key predictions of the SU(2) slave boson theory based on
the general resonating-valence-bond (RVB) picture.
\end{abstract}
 \maketitle

Since the discovery of the cuprate superconductors, 18 years have
elapsed without a consensus about its mechanism. Many exotic
features beyond the Bardeen-Cooper-Schrieffer theory have been
observed. One of them is the observation of a pseudogap in the
electron spectral function near the antinodal points ($\pi$,0) and
(0,$\pi$) at a temperature
$T^*>>T_c$\cite{Marshall,DingH,Pseudogap}. In a conventional BCS
superconductor, this gapping process occurs simultaneously with
the superconductivity at $T_c$. It has been heavily debated about
the relationship between the pseudogap and the superconductivity
in cuprates. One scenario assumes that the pseudogap $\Delta_p$
marks only a competing or coexisting order with the
superconductivity and it has nothing to do with the pairing
origin. However another picture, namely the Anderson's
resonating-valence-bond (RVB)\cite{RVB} model (and its
offspring)\cite{Prepair,PhaseString} predict that the spin-singlet
pairing in the RVB state (which causes the formation of the
pseudogap) may lend its pairing strength to the mobile electrons
and make them to naturally pair and then to condense at $T_c$.
According to this picture there should be a close relationship
between the pseudogap and the superconductivity.

In order to check whether this basic idea is correct, we need to
collect the information for both gaps, especially their doping
dependence. The pseudogap values $\Delta_p$ (or its corresponding
temperature $k_BT^* \sim \Delta_p$) and its doping dependence have
been determined through experiment\cite{Marshall,DingH,Pseudogap},
but it turns out to be a very difficult job to determine the
superconducting gap since both gaps entangle into each other in
the superconducting state. One exception is left, however, in the
small region of momentum space near the nodal point where the
pseudogap is generally assumed to be zero above $T_c$ and only the
superconducting gap opens below $T_c$. Therefore to detect the
weak gap information (or the gap slope
$v_\Delta=[d\Delta_s/d\phi]_{node}/\hbar k_F$) near nodal point in
the zero temperature limit becomes highly desired. Some previous
results using, for example, angle-resolved photo-emission
(ARPES)\cite{Mesot} or superfluid density seem to be in-conclusive
due to either temperature limitation (ARPES above 10 K) or
unexpected difficulty in analyzing the data (e.g., a so-called
Fermi liquid correction factor $\alpha_{FL}$ is inevitably
involved in analyzing the low temperature data of superfluid
density). In this paper, we report the evidence of a
proportionality between the nodal slope $v_\Delta$ of the
superconducting gap and the pseudogap temperature $T^*$.
Remarkably the maximum quasiparticle gap derived from $v_\Delta$
is also quite close to $T^*$. We also find that $T_c$ is
controlled by both the gap slope $v_\Delta$ and the size of the
Fermi arcs ($k_{arc}$) in the underdoped normal state. Both
observations are anticipated by the SU(2) slave boson
theory\cite{SU2} based on the general RVB picture.

We determine the properties of the nodal quasiparticles by
measuring the low temperature electronic specific heat. The
$La_{2-x}Sr_xCuO_4$ single crystals measured in this work were
prepared by travelling solvent floating-zone technique. Samples
with seven different doping concentrations p=0.063($T_c$=9K,
nominal x=0.063, post-annealed in $Ar$ gas at $800^\circ C$ for 48
hrs ), 0.069($T_c$=12K, as-grown sample with x=0.063), 0.075
($T_c$=15.6K, nominal x=0.07 and post-annealed in $O_2$ gas at
$750 ^\circ C$ for 12 hrs), 0.09 ($T_c$=24.4K, as grown, x=0.09),
0.11 ($T_c$=29.3K, as grown, x=0.11), 0.15 ($T_c$=36.1K, nominal
x=0.15), 0.22 ($T_c$=27.4K, nominal x=0.22) have been
investigated. The quality of our samples has been characterized by
x-ray diffraction, and $R(T)$ data showing a narrow transition
$\Delta T_c \leq $ 2 K. The samples have also been checked by AC
and DC magnetization showing also quite narrow transitions. The
full squares in Fig.3 represent the transition temperatures of our
samples. The heat capacity presented here was measured with the
relaxation method based on an Oxford cryogenic system
Maglab-EXA-12. In all measurements the magnetic field was applied
parallel to c-axis. As also observed by other groups for $La-214$
system, the anomalous upturn of $C/T$ due to the Schottky anomaly
of free spins is very weak. This avoids the complexity in the data
analysis. Details about the sample characterization, the specific
heat measurement, the residual linear term and extensive analysis
are reported in a recent paper\cite{WenPRB2004}.

It has been widely perceived that the pairing symmetry in the hole
doped cuprate superconductors is of d-wave with line nodes in the
gap function. In the mixed state, due to the presence of vortices,
Volovik \cite{Volovik} pointed out that supercurrents around a
vortex core lead to a Doppler shift to the quasi-particle
excitation spectrum. This will dominate the low energy excitation
and the specific heat (per mol) behaves
as\cite{Volovik,Kopnin1996} $C_{vol}=A\sqrt{H}$ with $A \propto
1/v_\Delta$. This square-root relation has been verified by many
measurements which were taken as evidence for d-wave symmetry, for
example by specific
heat\cite{Moler,Revaz,Wright,Phillips,Nohara,Chen,Hussey}, thermal
conductivity\cite{Taillefer}, tunnelling (to measure the Doppler
shift of the Andreev bound states)\cite{Greene,Deutscher},
NMR\cite{Slichter}, etc. In this way one can determine the nodal
slope ($v_{\Delta}$). Since the phonon part of the specific heat
is independent on the magnetic field, this allows to remove the
phonon contribution by subtracting the C/T at a certain field with
that at zero field, one has $\Delta\gamma=\Delta
C/T=[C(H)-C(0)]/T=C_{vol}/T-\alpha T$ with $\alpha$ the
coefficient for the quasiparticle excitations of a d-wave
superconductor at zero field ($C_e=\alpha T^2$). In the zero
temperature limit $\Delta \gamma=C_{vol}/T=A\sqrt{H}$ is
anticipated.

\begin{figure}
\includegraphics[width=8cm]{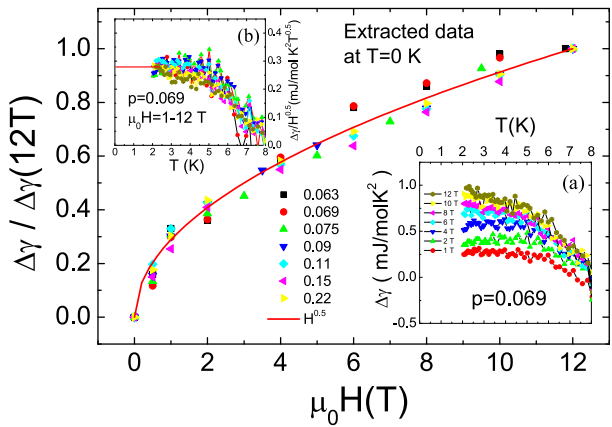}
\caption{Field dependence of $\Delta \gamma =[C(H)-C(0)]/T$
normalized by the data at about 12 T in zero temperature limit. It
is clear that Volovik's $\sqrt{H}$ relation describes the data
rather well for all samples. The inset(a) shows the typical
original data of $\Delta\gamma$ vs. $T$ for the underdoped sample
$p=0.069$. The inset (b) shows the same set of data as
$\Delta\gamma/\sqrt{H}$ vs. $T$. One can clearly see that in zero
temperature limit $\Delta\gamma / \sqrt{H}$ is a constant for all
fields implying the validity of the Volovik's relation
$\Delta\gamma=A\sqrt{H}$.} \label{fig1wen}
\end{figure}

In order to get $\Delta \gamma$ in the zero temperature limit, we
extrapolate the low temperature data of $C/T$ vs. $T^2$ (between
2K to 4K) to zero K. The data taken in this way and normalized at
12 T are presented in the main panel of Fig.1. It is clear that
the Volovik's $\sqrt{H}$ relation describes the data rather well
for all doping concentrations. This is to our surprise since it
has been questioned whether the Volovik relation is still obeyed
in the underdoped regime\cite{Nohara} especially when competing
orders are expected to appear\cite{SDW,DDW,AFSO5} and impurity
scattering is present. We attribute the success of using the
Volovik relation here to three reasons: (1) We use $\Delta
\gamma=[C_{H ||c}-C_{H=0}]/T$ instead of using $\Delta
\gamma=[C_{H ||c}-C_{H\perp c}]/T$. The latter may inevitably
involve the unknown DOS contributions from other kinds of vortices
(for example, Josephson vortices) when $H\perp C$. (2)The
contribution from a second competing order to $\Delta \gamma$ may
be small compared to the Volovik's term in the zero temperature
limit. (3)Single crystals today have much better quality leading
to much weaker impurity scattering. To have a self-consistent
check of the $\sqrt{H}$ relation found in the zero temperature
limit, we plot the data of $\Delta\gamma/\sqrt{H}$ vs. $T$ at
finite temperatures. A typical example for the very underdoped one
($p=0.069$) is shown in the inset(a) and (b) of Fig.1. One can see
that in the low temperature region the data
$\Delta\gamma/\sqrt{H}$ scale for all fields ranging from 1 T to
12 T, showing the nice relation $\Delta\gamma \propto \sqrt{H}$
for this sample in the zero temperature limit. From here one can
also determine the prefactor $A$ in $\Delta\gamma=A\sqrt{H}$ (here
for example, $A$=0.28 for p=0.069) and then compare backwards to
the value determined from the data shown in the main panel leading
to of course the same value. The same feature appears for all
other doping concentrations. For clarity they will not be shown
here.

\begin{figure}
\includegraphics[width=8cm]{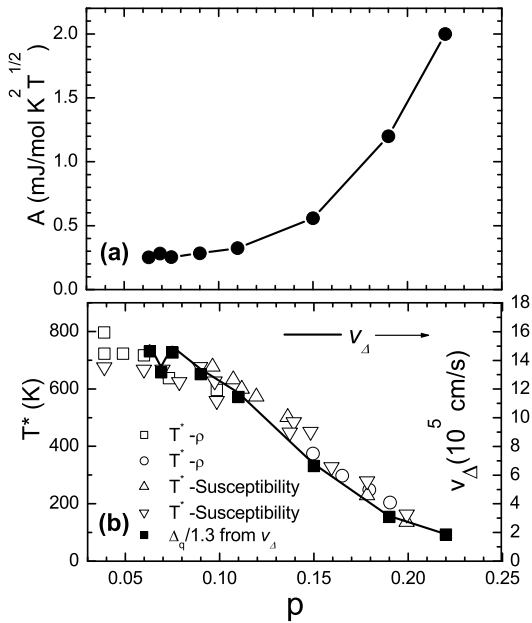}
\caption {(a) Doping dependence of the pre-factor $A$ determined
in present work (full circles). Here the point at p=0.19 was
adopted from the work by Nohara et al. on a single
crystal\cite{Nohara}. (b) Doping dependence of the pseudogap
temperature $T^*$ (open symbols) summarized in Ref.3 and our data
$v_\Delta$ (solid line). $T^*$-susceptibility refers to the
pseudogap temperature determined from the maxima in the static
susceptibility, and $T^*-\rho$ to the temperature at which there
is a slope change in the DC resistivity. Above $T^*-\rho$ the
resistivity has a linear temperature dependence. The full squares
represent the calculated virtue maximum quasi-particle gap
$\Delta_q$ derived from $v_\Delta$ without any adjusting
parameters. Surprisingly both set of data are close to each other
($\Delta_q \sim 1.3k_BT^*$) although they are determined in
totally different experiments. This result implies a close
relationship between the pseudogap $\Delta_p$ and the
superconducting gap slope $v_\Delta$. } \label{fig2wen}
\end{figure}

\begin{figure}
\includegraphics[width=8cm]{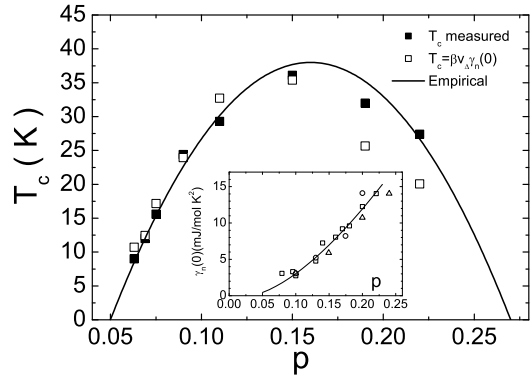}
\caption{ Doping dependence of the really measured superconducting
transition temperature $T_c$ (full squares) and that calculated by
$T_c=\beta v_\Delta \gamma_n(0)$ (open squares) with $\beta=0.7445
K^3 mol s/J m$. The solid line represents the empirical relation
$T_c/T_c^{max}=1-82.6(p-0.16)^2$ with $T_c^{max}=38 K$. The inset
shows the doping dependence of $\gamma_n$ derived from specific
heat (open squares\cite{Momono}),ARPES (open circles\cite{Ino}),
and Knight shift (up-triangles\cite{Ohsugi}). The solid line is a
fit to the data with $\gamma_n=\zeta(p-p_c)^\eta$.}
\label{fig3wen}
\end{figure}

This successful scaling of $\Delta \gamma$ vs. $\sqrt{H}$ makes it
possible to derive the pre-factor $A$, and one can further
determine the gap slope $v_\Delta$. Fig.2(a) shows the doping
dependence of the pre-factor $A$. For a typical d-wave
superconductor, by calculating the excitation spectrum near the
nodes, it was shown that\cite{Hussey}
\begin{equation}
\label{deltap}
 A=\alpha_p\frac{4k_B^2}{3\hbar l_c}\sqrt{\frac{\pi}{\Phi_0}}\frac{nV_{mol}}{v_\Delta}
\end{equation}
here $l_c$ = 13.28 \AA is the c-axis lattice constant, $V_{mol}$
=58 $cm^3$ (the volume per mol), $\alpha_p$ a dimensionless
constant taking 0.5 (0.465) for a square (triangle) vortex
lattice, n=2 (the number of Cu-O plane in one unit cell), $\Phi_0$
the flux quanta. The $v_\Delta$ has then been calculated without
any adjusting parameter (taking $\alpha_p$=0.465) and shown in
Fig.2(b). It is remarkable that $v_\Delta$ has a very similar
doping dependence as the pseudogap temperature $T^*$, indicating
that $v_\Delta\propto T^*\propto \Delta_p$. If converting the data
$v_\Delta$ into the virtue maximum quasiparticle gap
($\Delta_q$)\cite{Hussey} via $v_\Delta=2\Delta_q a $ with
$a=3.8$\AA (the in-plane lattice constant), surprisingly the
resultant $\Delta_q$ value [shown by the filled squares in
Fig.2(b)] is quite close to $T^*$ ($\Delta_q \sim 1.3 k_B T^*$).
It is important to stress that this result is obtained without any
adjusting parameters. Counting the uncertainties in determining
$T^*$ and the value of $\alpha_p$, this relation is remarkable
since $\Delta_q$ and $T^*$ are determined in totally different
experiments. Because $v_\Delta$ and $\Delta_q$ reflect mainly the
information near nodes which is predominantly contributed by the
superconducting gap, above discovery, i.e., $v_\Delta\propto
T^*\propto \Delta_p$ (or $\Delta_q \sim k_BT^*$) strongly suggests
a close relationship between the superconductivity and the
pseudogap. A similar conclusion was drawn in underdoped
$YBa_2Cu_3O_y$ by analyzing the low temperature thermal
conductivity\cite{Taillefer2}. Since the pseudogap is supposed to
be caused by the formation of the RVB state\cite{RVB}, our results
here point to a fact that the RVB singlet pairing may be one of
the unavoidable ingredients for superconductivity.

In the following we will investigate what determines $T_c$.
Bearing the doping dependence of $v_\Delta$ in mind, it is easy to
understand that $v_\Delta \hbar k_F$ should not be a good estimate
of the superconducting gap for underdoped samples. The basic
reason is that the normal-state Fermi surfaces are small arcs of
length $k_{arc}$ near the nodal points. The superconducting
transition only gaps the Fermi arc. So the effective
superconducting gap should be estimated as $\Delta_s \sim
\frac{1}{2}v_\Delta \hbar k_{arc}$. From the normal state
electronic specific heat $C_{ele}=\gamma_nT$,
$\gamma_n \sim 2nk_B^2k_{arc}V_{mol}/3\hbar v_Fl_c$,
we can obtain $k_{arc}$. Assuming $\Delta_s \sim k_B T_c$, we find
\begin{equation}
\label{Tc}
 T_c=\alpha_s^{-1} \frac{3\hbar^2 v_F l_c \gamma_n v_\Delta}{4n k_B^3
 V_{mol}}=\beta\gamma_nv_\Delta
\end{equation}
where $\alpha_s$ is a dimensionless constant.
$v_F$ is the nodal Fermi velocity normal the Fermi surface.
The value of $\gamma_n(0)$ can be estimated from specific
heat\cite{Momono}, or indirectly by ARPES\cite{Ino} or
NMR\cite{Ohsugi}. Here we take the values for $\gamma_n(0)$
summarized by Matsuzaki et al.\cite{Momono} and fit it (in unit of
$mJ/mol K^2$) with a formula $\gamma_n=\zeta(p-p_c)^\eta$ yielding
$\zeta=182.6,p_c=0.03, \eta=1.54$. In Fig.3 we present the doping
dependence of the really measured $T_c$ (filled squares) and the
calculated value (open squares) by eq.(2) with $\beta=0.7445
K^3mols/J m$. In underdoped region, the really measured and
calculated $T_c$ values coincide rather well implying the validity
of eq(2). In the overdoped region, $\gamma_n$ will gradually
become doping independent, therefore one expects $T_c\propto
v_\Delta$. Using $\beta=0.7445 K^3mols/J m$ and taking
%$\alpha_s=0.434$
$\alpha_s=13.8$, we get $v_F=2.73 \times 10^7 cm/s$ which is the
so-called universal nodal velocity determined by
ARPES\cite{ZhouXJ}. So the energy scale of the superconductivity
is not given by $v_\Delta \hbar k_F\sim \Delta_p$, but by
$\frac{1}{2}v_\Delta \hbar k_{arc}$ or more precisely by eq.(2).

To have a framework about the experimental results, in the
following, we will review one particular explanation based on the
slave-boson approach. Within the SU(2) slave-boson
theory\cite{SU2}, the pseudogap metallic state is viewed as a
doped algebraic spin liquid (ASL)\cite{ASL}. A doped ASL is
described by spinons (neutral spin-1/2 Dirac fermions) and holons
(spinless charge-e boson) coupled to a U(1) gauge field. Due to
the attraction between the spinons and the holons caused by the
U(1) gauge field, a spinon and a holon recombine into an electron
at low energies\cite{SU2,ASL}. Due to the spin-charge
recombination, the pseudogap metallic state is described by
electron-like quasiparticles at low energy. Since the binding
between the spinon and the holon is weak, the large pseudogap near
the anti-nodal points ($\pi$,0) and (0, $\pi$) is not affected. So
the Fermi surface of the recombined electrons cannot form a large
closed loop. A simple theoretical calculation\cite{SU2} suggests
that the Fermi surface of the recombined electrons forms four
small arcs near the nodal points ($\pm \pi/2,\pm \pi/2$). Thus the
SU(2) slave boson theory\cite{SU2} contains two important features: the
pseudogap due to spin singlet pairing and the Fermi arcs due to
the spin-charge recombination\cite{ASL}. The superconductivity
arises from the condensation of the quasiparticles on the arcs,
thus one expects that $T_c$ is proportional to the gap on the
Fermi arc: $k_BT_c \sim \frac{1}{2}v_\Delta \hbar k_{arc}$,
instead of the pseudogap $\Delta_p$ near the anti-nodal points.
Meanwhile, since the spin pairing is responsible for both the
pseudogap $\Delta_p$ near the anti-nodal points and the
superconducing gap slope $v_\Delta$, one finds that
$v_\Delta$ is propotional to $T^*(\propto \Delta_p)$ or
$\Delta_q\sim k_BT^*$ in the simplest version of
the slave-boson theory.\cite{SU2} These are exactly what we found in the
experiment.

In summary, the Volovik's relation of the d-wave pairing symmetry
has been well demonstrated by low temperature specific heat in
wide doping regime in $La_{2-x}Sr_xCuO_4$. From here the nodal
slope $v_\Delta$ of the superconducting gap is derived and is
found to follow the same doping dependence of the pseudogap
$\Delta_p$. Meanwhile it is found that the superconducting
transition temperature $T_c$ is controlled by $v_\Delta
\gamma_n(0)$ instead of $v_\Delta$. Both observations are
consistent with the SU(2) slave boson theory based on the general
RVB picture.

This work is supported by the National Science Foundation of China
, the Ministry of Science and Technology of China, the Knowledge
Innovation Project of Chinese Academy of Sciences. XGW is
supported by NSF Grant No.DMR--04--33632, NSF-MRSEC Grant No.
DMR--02--13282, and NFSC no. 10228408. We thank Yoichi Ando and
his group (CRIEP, Komae, Tokyo, Japan) for providing us some nice
single crystals. We are grateful for fruitful discussions with Z.
Y. Weng, T. Xiang and Q. H. Wang.

Correspondence should be addressed to:\\
 hhwen@aphy.iphy.ac.cn or wen@dao.mit.edu.


\begin{thebibliography}{00}
\bibitem{Marshall}D. S. Marshall, et al., Phys. Rev. Lett.{\bf76}, 4841(1996).
\bibitem{DingH}H. Ding, et al., Nature {\bf382}, 51-54 (1996).
\bibitem{Pseudogap}For a review see, for example, T. Timusk, B. Statt,
Rep. Prog. Phys.{\bf62},61(1999).
\bibitem{RVB}For a recent review on RVB picture, see
Patrick A. Lee, Naoto Nagaosa, Xiao-Gang Wen, cond-mat/0410445; P.
W. Anderson, P. A. Lee, M. Randeria, T. M. Rice, N. Trivedi, and
F. C. Zhang, J. Phys. Condens. Matter {\bf16}, R755 (2004); P. W.
Anderson et al., Phys. Rev. Lett.{\bf58}, 2790(1987); P. W.
Anderson, Science{\bf235}, 1196(1987); Z. Y. Weng, D. N. Sheng, C.
S. Ting, Phys. Rev. Lett.{\bf80}, 5401 (1998).
\bibitem{Prepair} S. A. Kivelson and V. J. Emery,
Nature {\bf374}, 434(1995).
\bibitem{PhaseString}Z. Y. Weng, et al., Phys. Rev. B{\bf55}, 3894(1997).
\bibitem{Mesot}J. Mesot, et al. Phys. Rev. Lett.{\bf83}, 840(1999).
\bibitem{SU2}X. G. Wen,  and P. A. Lee, Phys. Rev. Lett. {\bf76}, 503(1996);
P. A. Lee, and X.-G. Wen, Phys. Rev. Lett.{\bf78}, 4111(1997);
X.-G. Wen, and P. A. Lee, Phys. Rev. Lett.{\bf80}, 2193(1998).
\bibitem{WenPRB2004}H. H. Wen, et al., Phys. Rev. B{\bf70}, 214505(2004).
\bibitem{Volovik} G.E. Volovik, JETP Lett. {\bf 58}, 469 (1993);
ibid {\bf65}, 491 (1997).
\bibitem{Kopnin1996} N. B. Kopnin and G. E.
Volovik, JETP Lett. {\bf 64}, 690 (1996).
\bibitem{Moler} K. A. Moler, et al., Phys. Rev. Lett.
{\bf73}, 2744 (1994). K. A. Moler, et al., Phys. Rev. B{bf 55},
12753 (1997).
\bibitem{Revaz} B. Revaz, et al., Phys. Rev. Lett. {\bf80}, 3364 (1998).
\bibitem{Wright} D. A. Wright, et al., Phys. Rev. Lett.{\bf82}, 1550 (1999).
\bibitem{Phillips} N. E. Phillips, et al., Physica B {\bf259-261},
546 (1999).
\bibitem{Nohara}M. Nohara, et al., J. Phys. Soc. Jpn. {\bf69},
1602 (2000).
\bibitem{Chen}S. J. Chen, et al., Phys. Rev. B {\bf58}, R14753 (1998).
\bibitem{Hussey}For a review on the low energy quasiparticles, see N. E. Hussey, Advances in Physics {\bf51}, 1685 (2002).
\bibitem{Taillefer}M. Chiao, et al., Phys. Rev. Lett.{\bf 82}, 2943(1999).
\bibitem{Greene}M. Aprili, E. Badica, L. H. Greene, Phys. Rev. Lett.{\bf 83}, 4630(1999).
\bibitem{Deutscher}R. Beck, et al., Phys. Rev. {\bf B 69}, 144506(2004).
\bibitem{Slichter}N. J. Curro, et al., Phys. Rev.{\bf B 62}, 3473(2000).
\bibitem{SDW} E. Demler, S. Sachdev and Y. Zhang, Phys. Rev. Lett.{\bf87},
067202(2001); Y. Zhang, E. Demler and S. Sachdev, Phys. Rev.{\bf
B66}, 094501(2002).
\bibitem{DDW}
Sudip Chakravarty, R. B. Laughlin, Dirk K. Morr, and Chetan Nayak,
Phys. Rev. {\bf B63}, 094503 (2001).
%S. Chakravarty, and H.-Y. Kee, Phys. Rev. {\bf B61},
%14821(2000). ibid {\bf62}, 4880(2000); {\bf62}, R6135(2000).
\bibitem{AFSO5} S. C. Zhang, Science {\bf275}, 1089(1997).
\bibitem{Taillefer2}M. Sutherland, et al., Phys. Rev. B {\bf67}, 174520(2003).
\bibitem{Momono} T. Matsuzaki, N. Momono, M. Oda, M. Ido, J. Phys. Soc. Japn. {\bf73}, 2232(2004).
\bibitem{Ino}A. Ino, et al., Phys. Rev. Lett.{\bf81}, 2124(1998).
\bibitem{Ohsugi}S. Ohsugi, et al., J. Phys. Soc. Jpn.{\bf63}, 700(1994).
\bibitem{ZhouXJ}X. J. Zhou, et al., Nature {\bf423}, 398(2003).
\bibitem{ASL}W. Rantner, X. G. Wen, Phys. Rev. B{\bf66}, 144501(2002).

% \bibitem{label}
% Text of bibliographic item

% notes:
% \bibitem{label} \note

% subbibitems:
% \begin{subbibitems}{label}
% \bibitem{label1}
% \bibitem{label2}
% If there is a note, it should come last:
% \bibitem{label3} \note
% \end{subbibitems}


\end{thebibliography}
\end{document}